\documentclass[twocolumn,amssymb,showpacs,amsmath,nobibnotes,nofootinbib,aps,prb]{revtex4}

\usepackage{graphicx}
\usepackage{dcolumn}
\usepackage{bm}

\begin{document}

\preprint{APS/123-QED}

\title{Intraplanar Magnetic Excitations in Na$_{\frac{1}{2}}$CoO$_2$:\\ 
An Inelastic Neutron Study}

\author{J. Wooldridge$^1$}
 \email{J.Wooldridge@warwick.ac.uk}

\author{G. Balakrishnan$^1$}
\author{D. M$^c$K. Paul$^1$}%
\author{C. Frost$^2$} 
\author{P. Bourges$^3$}
\author{M. R. Lees$^1$}

\affiliation{
$^1$Department of Physics, Univeristy of Warwick, Coventry, CV4 7AL, UK
}

\affiliation{
$^2$ISIS, Rutherford Appleton Laboratory, Chiltern, Didcot, OX11 0QZ, UK
}

\affiliation{
$^3$Laboratoire L$\acute{\textmd{e}}$on Brillouin (CEA-CNRS) CEA-Saclay, 91191 Gif-sur-Yvette Cedex, FRANCE
}

\date{\today}

\begin{abstract}
Inelastic neutron scattering measurements mapping the in-plane magnetic interactions of Na$_{\frac{1}{2}}$CoO$_2$ reveal dispersive excitations at points above an energy gap E$_{g}$ = 11.5(5) meV at the superstructural Bragg reflections. The excitations are highly damped, broadening with increasing energy, and disappear at $\hbar \omega \approx$35 meV, a strong indication that the magnetism is itinerant. Tilting into the \textit{ac} plane reduces the value of E$_{g}$ by 25\%, suggesting that the dispersion along \textit{c} is significant and the magnetic correlations are three-dimensional, as seen at the higher doping levels.  
\end{abstract}

\maketitle

The discovery of superconductivity in hydrated sodium cobaltate in 2003~\cite{takada} has generated a large volume of study in a relatively short space of time. The anhydrous parent phase is interesting in its own right; the system, at x$\geq$0.65, is a Curie-Weiss metal with magnetic ordering below 22 K and the signature of magnetism, assigned to a spin density wave (SDW), depends sensivitely on the level of sodium doping via its influence on the Co valence state. Recent neutron investigations of the SDW phase~\cite{helme,bayrakci} demonstrated that the magnetic ordering can be ascribed to an A-type antiferromagnetic (A-AFM) structure. In this system, however, only $1-x$ of the Co sites carry spin, so that the simple Hamiltonian used to describe the excitation modes would only be valid if phase segregation into magnetic and nonmagnetic regions is assumed. This raises the question of whether the system really can be characterised by a model with localised spin ordering or whether an itinerant model with small charge disproportionation is more appropriate.
A new regime emerges at x = $\frac{1}{2}$, with different magnetic ordering. At T$_M$ = 88 K, long-ranged static AFM ordering occurs which has been clearly identified in neutron diffraction experiments~\cite{gasp}. At T$_{MI}$ = 53 K, the material undergoes a metal to insulator transition (MIT). Further to this, there exists long-ranged ordering of the Na cations~\cite{huang} into quasi-onedimensional chains along \textit{a}, enabling the crystal structure to be reclassified into an orthorhombic supercell of dimensions ($2a\times\sqrt{3}a\times c$). In this crystal setting, there are now two Co sites. This led to speculation that the above transitions were the results of charge ordering (CO) in the CoO$_2$ planes, probably due to the influence of charge modulations in the Na layer. More recently, however, the notion of a charge ordered Mott-like transition has been disputed. NMR measurements~\cite{bobroff} did not deduce any large difference in charge state between the two sites. Crystal structure determination by powder neutron diffraction~\cite{williams} also failed to find a significant difference between the Co-O hybrization on the two sites.
A clear picture of the role of spin and orbital ordering in Na$_{\frac{1}{2}}$CoO$_2$ has yet to be defined. Here we present the first measurements of inelastic neutron scattering on this system to probe the nature of the interactions in the \textit{ab} planes. A dispersive mode, magnetic in origin, is observed; the high damping and large velocity of the excitation indicates the possibility that the AFM ground state is itinerant. Preliminary investigations indicate that the system is also highly dispersive along \textit{c}.
\par

The samples were prepared by the floating zone method as described previously~\cite{me}. To confirm the stoichiometry of each sample, given that the magnetic transitions at T$_{MI}$ and T$_M$ occur only at x = $\frac{1}{2}$, magnetic susceptibility ($\chi$) measurements were carried out in a Quantum Design MPMS-5S superconducting quantum interference device (SQUID) magnetometer on a range of samples; those selected for the experiments below had $\chi$ similar to previously reported measurements~\cite{gasp,bobroff,huang}. The results indicated the sample was free from magnetic impurities (CoO or Co$_3$O$_4$). The heat capacity (C) measurements were carried out by a two-tau relaxation method in a Quantum Design PPMS. Inelastic neutron scattering measurements were taken on the MAPS spectrometer at ISIS (RAL) and the triple axis spectrometer (TAS) 2T1 at the Laboratoire L$\acute{\textmd{e}}$on Brillouin (LLB) (CEA-CNRS). A mosaic of six crystals, each of mass $\sim$0.5g, was used for the experiment at ISIS. The samples were loaded into a CCR and cooled to 10 K. An incident neutron energy of 100 meV was selected and long counting times (typically 48 hours) were adopted to build up sufficient statistics in (\textbf{Q},E) space over the range of interest. Just one of these crystals was selected for the triple axis measurement in order to reduce the overall mosaic spread and optimise resolution. The crystal was cooled to 2.8 K in an orange cryostat with the \textit{c} axis aligned vertically so that the Co moments were in the scattering plane. 

\begin{figure}[t]
\includegraphics[width = 7.5 cm]{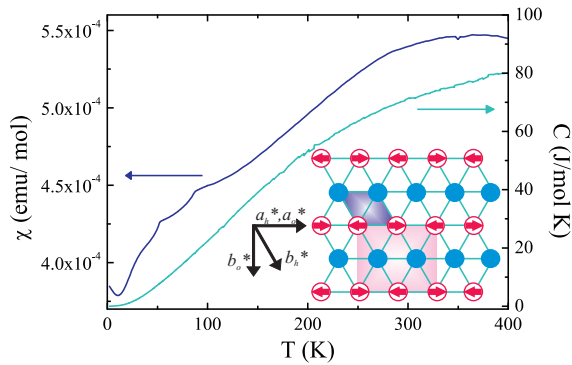}
\caption{\label{mcdata} Magnetic susceptibility (H$_{\textit{ab}}$ = 5 T)and heat capacity of Na$_{\frac{1}{2}}$CoO$_2$. The two transitions at T$_M$ and T$_{MI}$ are clearly visible in the magnetisation. On the contrary, the heat capacity data is featureless. The magnetic structure, first proposed by Choy et al.~\cite{choy} is presented in the inset, both the hexagonal (shaded blue) and orthorhombic (pink) unit cells are shown.}
\end{figure}

\par
Magnetic susceptibility and heat capacity measurements are shown in figure~\ref{mcdata}. Both the magnetic ordering and the metal-insulator transitions are clearly visible as shoulders in the magnetisation. It is interesting to note that $\chi$ does not exhibit the Curie-Weiss-like character as seen in the SDW material: there is no signature of charge locality from the macroscopic measurements. 
The magnetic structure as measured by Ga$\check{\textmd{s}}$parovi$\acute{\textmd{c}}$ et al.~\cite{gasp} is presented in the inset. The symmetry breaking imposed by the orthorhombic supercell produces two separate Co sites, one of which is magnetically ordered (Co$^{4+}$ S=$\frac{1}{2}$) and the other nonmagnetic (Co$^{3+}$ S=0). In terms of the new unit cell, the system can be thought of as a simple G-type antiferromagnet with a magnetic propagation vector of ($1\mathit{\frac{1}{2}}1$). The dynamics of the magnetism may be more complicated than suggested by this localised moment ordering picture; fluctuations on the nonordered rows of Co may affect the spin wave dispersion. The fact that the MIT occurs 35 K below T$_M$ would suggest the transport carriers reside mainly on the second Co site. The opening of a gap on the part of the Fermi surface (FS) related to band containing these carriers would readily explain the onset of the insulating transition.

\par
\begin{figure}[b]
\includegraphics[width = 7.5 cm]{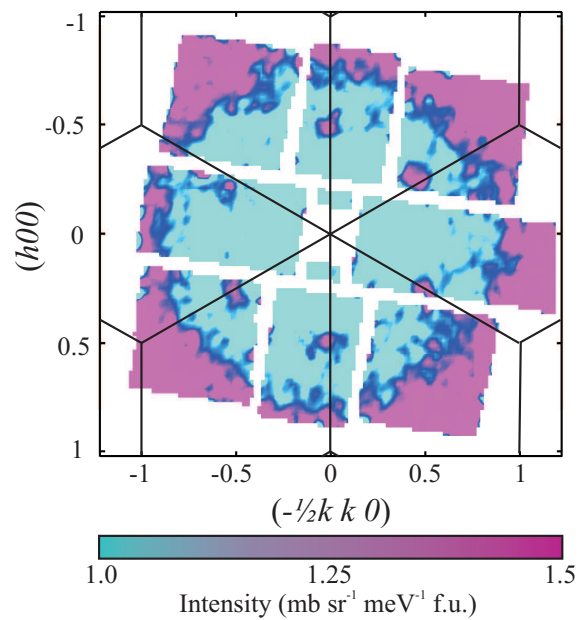}
\caption{\label{mapsdata} Inelastic scattering in the ($hk0$) plane summed over energy transfers of 10-20 meV. Scattering is visible at points corresponding to [($100$)]($\frac{1}{2}00$) in the [orthorhombic]hexagonal [super-]cell, although twinning reproduces the hexagonal symmetry. An overlay of the hexagonal reciprocal lattice is shown.}
\end{figure}
\par
In order to confirm the nature of the magnetic ordering proposed by Ga$\check{\textmd{s}}$parovi$\acute{\textmd{c}}$ et al.~\cite{gasp} the crystals were first rotated by 30$^\circ$ to the incoming neutron beam in order to mimic the experimental setup as used by Boothroyd et al.~\cite{boothroyd}, where in-plane ferromagnetic (FM) ordering for an x = 0.7 crystal was visible as scattering around the origin, corresponding to a zero 2D magnetic ordering vector. No such scattering exists at this sodium level confirming the absence of FM coupling. The crystal was then rotated back to have k$_i$ $\|$ \textit{c}. Inelastic scattering intensity is clearly visible (figure~\ref{mapsdata}) at points above ($\mathit{\frac{1}{2}}00$) in the parent hexagonal lattice. All six structural Bragg peaks are also visible in the elastic channel, which is at odds with a true long-ranged orthorhomic superstructure, where there should only be two. However, the integrated intensity of pairs of peaks diametrically opposite each other match, suggesting that the crystals exhibit twinning (rather tripling) in the basal plane and are not just aligned along different equivalent direction choices of ($100$).
S(\textbf{Q},$\hbar\omega$) maps were produced by slicing the data across different energy transfer levels. The small size of the sample resulted in a low signal to background ratio; the background itself is a product of multiple and instrumental scattering and increases with $\left| \textbf{Q} \right|$. The energy resolution is dependent on the opening time of chopper used to monochromate the beam; this was relaxed to collect good statistics but ensured that elastic scattering extended up to 10 meV.  As a result the background obscures the magnetic scattering at $\left| \textbf{Q} \right| \geq$0.75. Magnetic excitations should be visible as a cone of scattered intensity above the magnetic zone centre. There are no in-plane magnetic Bragg reflections~\cite{gasp} and so, to further investigate the origin of the scattering seen here, measurements were completed at the TAS 2T1 at LLB. 
\par
\begin{figure}[t]
\includegraphics[width = 7.5 cm]{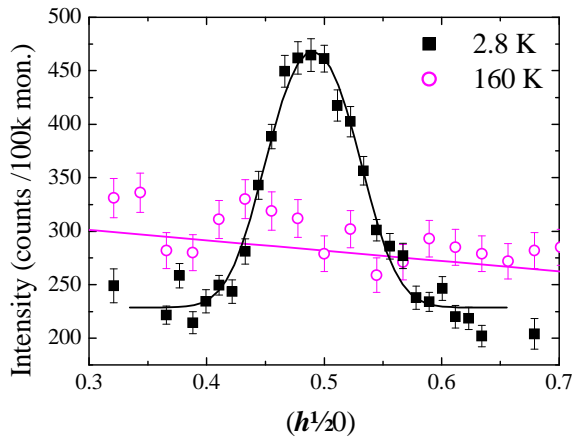}
\caption{\label{dept} Constant energy scans through ($\mathit{\frac{1}{2}\frac{1}{2}}0$) at $\hbar \omega$ = 15 meV. The superstructural Bragg peak at this \textbf{Q}, and therefore the Na ordering, persists to high temperatures, whilst magnetic Bragg peaks at ($\mathit{\frac{1}{2}\frac{1}{2}}l_{odd}$) disappear at T$_M$~\cite{gasp}. The inelastic signal has disappeared at twice the ordering temperature T$_M$, with an increased background corresponding to an enhancement of the incoherent and multi-phonon scattering.}
\end{figure}
\begin{figure}[b]
\includegraphics[width = 7.5 cm]{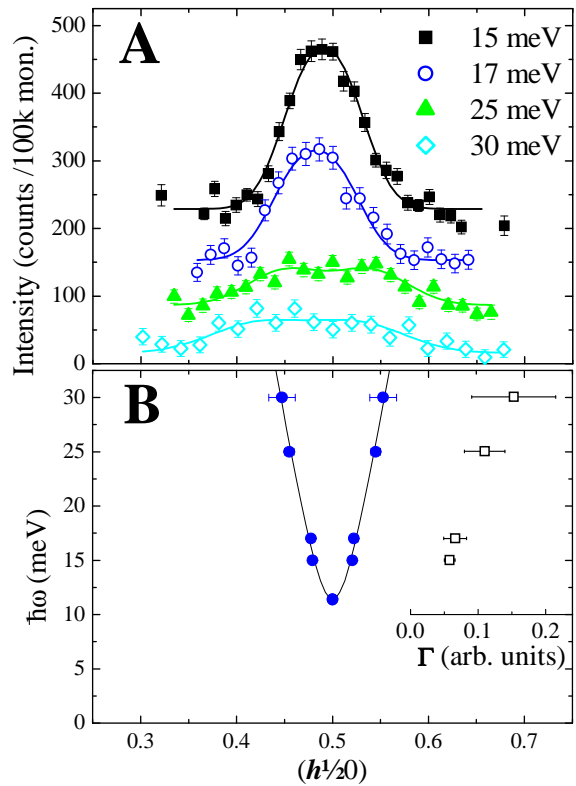}
\caption{\label{constE} A: Constant energy transfer scans at \textbf{Q}=($\mathit{\frac{1}{2}\frac{1}{2}}0$) taken along ($100$) at 2.8 K. The datasets have been offset by -150, -200 and -250 counts respectively for increasing energy transfer cuts. B: The measured spin wave dispersion along ($h\mathit{\frac{1}{2}}0$) (the variation of the spin wave damping $\Gamma$ with energy is shown in the inset); the excitation appears at $\sim$11.5 meV.}
\end{figure}

At a temperature of 2.8 K, a clear peak is identifiable in both the inelastic and elastic channels (a non-magnetic superstructural peak resulting from Na ordering~\cite{huang}) at \textbf{Q}=($\mathit{\frac{1}{2}}00$). In order to access high enough energy transfers to examine the dispersion, it was necessary to move to the next Brillioun zone. The magnitude of the change in intensity between the two sites is comparable to that expected from the decrease in the magnetic form factor of Co$^{4+}$ with the increase in $\left|\textbf{Q}\right|$. A further indication that the signal is a result of 3D magnetic ordering is that the peak is destroyed by increasing the temperature to $\sim$2T$_M$ (figure~\ref{dept}).
Similar sets of \textbf{Q}-scans were made at increasing energy transfers and are shown in figure~\ref{constE}. The excitation is gapped, with no scattered intensity visible below $\sim$11.5 meV. The peaks are also overly damped, increasing in width with energy, whilst the overall dispersion is very steep. The data can be fitted to a linear dispersion $\hbar \omega = \sqrt{\Delta^2 + \left( D\textbf{Q} \right)^2}$ where $\Delta$ is the value of the energy gap and $D$ the spin wave stiffness constant, more commonly referred to as the spin wave velocity in AFM systems. A fit to the data gives $\Delta$ = 11.5(5) meV and $D$ = 510(20) meV $\textmd{\AA}$ and is plotted as the solid line in panel B of figure~\ref{constE}.

\par

\begin{figure}[t]
\includegraphics[width = 7.5 cm]{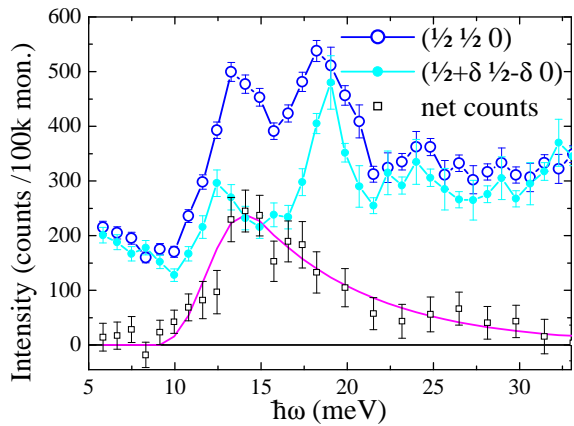}
\caption{\label{ecut} Energy scans at ($\mathit{\frac{1}{2}\frac{1}{2}}0$). A background was taken at a position of ($\mathit{\frac{1}{2}}+\delta \, \mathit{\frac{1}{2}}-\delta \, 0$) where $\delta\sim$0.1 was chosen so that the spectrometer was aligned with the same 2$\theta$ positions as the on-peak measurements. The subtracted data (black open squares) is plotted; the line corresponds to the convolution of the experimental resolution ellipsoid and the intensity expected for an ideal gapped system (E$_g$ = 11.5 meV).}
\end{figure}

To examine the excitation gap further, scans at constant \textbf{Q} were made with increasing energy transfer. The gap was observed at both ($\mathit{\frac{1}{2}}00$) and ($\mathit{\frac{1}{2}\frac{1}{2}}0$) (figure~\ref{ecut}) at E$_g$= 11.5 meV. The onset of scattering is sharp, and decays with increasing $\hbar\omega$. As seen at the higher doping levels, there exists an optical phonon (OP) at $\sim$20 meV which dominates the signal in the Q-scans at that $\hbar\omega$. Since OP modes are relatively flat, it was possible to pick up the signal from the magnetic excitation once more at higher $\hbar\omega$. In order to correct for the additional phonon component to the scattering, another energy scan was taken at an offset position from the excitation at (0.594 0.4 0), the exact values of ($hk0$) being chosen to set the analyzer at the same 2$\theta$ position as the previous scan. The difference curve then removes the incoherent[phonon] scattering below[above] the gap.
Finally, the sample was tilted slightly into the ($100$)-($001$) scattering plane and the same energy scan repeated for Q=($\mathit{\frac{1}{2}}0\mathit{\frac{1}{2}}$) (not shown). The value of E$_g$ is shifted to $\sim$9 meV suggesting that the dispersion along \textit{c} is significant, a feature in accord with the 3D magnetic correlations reported for $x >$0.7~\cite{johannes}. \textbf{Q} = ($\mathit{\frac{1}{2} \frac{1}{2}}0$) is not a magnetic zone centre, no magnetic scattering exists at this point. In order to determine whether the excitation includes a spin wave gap, information of the dispersion along \textit{c} is necessary.  
\par

These data exhibit features that are common to intinerant magnetic systems such as Cr~\cite{cr} or Mn$_3$Si~\cite{mn3si}, in which the highly damped dispersion, at the position of a magnetic Bragg peak, is seen to vary linearly in \textbf{Q} up to energies of $\sim$2 k$_B$T$_N$ and then rise vertically in a ``chimney'' of scattering. The dispersion here, however, remains linear in \textbf{Q} until the signal disappears at $\sim$4 k$_B$T$_N$. The most obvious method for determining the proportion of localised to itinerant spins in a FM material is to calculate the ratio of the high temperature paramagnetic moment and the low temperature saturated moment, equal to one in completely localised magnetic systems and greater than one for itinerant materials. This method is not applicable here; $\chi$ is not Curie-Weiss-like and the material has AFM ordering. For a localised spin system, the energy of the excitation at the zone boundary should be of the order of k$_B$T$_M$. An extroplation of the data collected in this study predicts these two energies to be different by an order of magnitude. If the system exhibits low dimensionality, the energy scale characterising the excitation will relate to the strength of interactions along the spin chains, thus underestimating k$_B$T$_M$. However the loss of the excitation into a broadened continuum of decay modes, as witnessed here, indicates strongly that the material is itinerant in character.
\par
In the light of these results, the previously proposed model of a charge ordered ground state should be reconsidered. No evidence of either the T$_M$ or T$_{MI}$ transition can be seen in the specific heat. Recent NMR studies~\cite{bobroff} have suggested that both transitions could be attributed to spin density waves i.e. FS nesting effects, in which case T$_{MI}$ would involve only modest modifications to the spin order, producing little measureable magnetic entropy. Alternatively, as mentioned above, the magnetism in Na$_{\frac{1}{2}}$CoO$_2$ may exhibit low dimensionality; in-plane correlations build slowly with decreasing temperature before locking in-phase along \textit{c} at T$_{M}$. The phase diagram of sodium cobaltate is characterised by the effects of strong electron correlations. At x$>$0.5, some of the Co states may be localised by these interactions causing the Curie-Weiss-like magnetic behaviour and spin wave excitations, whilst the material retains metallic conductivity. Below half doping the electron correlation effects are weaker and the system is Pauli paramagnetic. Therefore there must exist a cross-over between these two regimes, a special state realised at x=0.5 in which the magnetism is largely descibed by the itinerant electron picture but electron correlation effects play a role in the formation of the MIT. It should be noted that the insulating transition is anisotropic; the resisitivity at low temperatures along \textit{c} is 25 $\Omega$cm and just 15 m$\Omega$cm along \textit{ab}~\cite{wang2}, a value comparable to that of the ``metallic'' SDW system~\cite{me}, and whilst Na$_{\frac{1}{2}}$CoO$_{2}$ is considered a low temperature insulator in dc, optical conductivity measurements~\cite{wang3} reveal the MIT is only visible in the low frequency data. The mechanism that drives the MIT is clearly not yet understood and further investigation into the electronic band structure at this doping level is needed. 

\par
In summary, the intraplanar magnetic excitations whose zone centres coexist with structural Bragg reflections have been studied. The excitation has a gap of 11.5(5) meV. The localized spin model does not provide physically realistic values for the magnon velocity or the energy transfer at the zone boundary and is not consistent with the broadening and diminishing scattering more reminiscent to the transformation between spin wave and Stoner excitations as seen in the itinerant magnets. The excitation modes measured in this study indicate that the system is at least partially itinerant and that a model containing zero spin Co$^{3+}$ and spin-half Co$^{4+}$ is not appropriate. 

\par
We thank Julie Staunton and Jim Hague for insightful discussions and acknowledge financial support from the EPSRC. Both neutron scattering experiments were funded within th EU NMI3 access program.

\bibliography{MAPSLLB}

\end{document}